\documentclass[showpacs,preprintnumbers,two column]{revtex4}
\usepackage{amsmath}
\usepackage{graphicx,color}
\usepackage{dcolumn}
\usepackage{bm}
\usepackage{ulem}

\begin{document}
\title{Application of polaron picture in the two-qubit quantum Rabi model}
\author{Xi-Mei Sun$^{1}$, Lei Cong$^2$, Hans-Peter Eckle$^{3}$, Zu-Jian Ying$^{1}$ and Hong-Gang Luo$^{1,4}$}

\address{
$^{1}$Center for Interdisciplinary Studies and Key Laboratory for Magnetism and Magnetic Materials of the MoE,
Lanzhou University, Lanzhou 730000, China\\
$^{2}$International Center of Quantum Artificial Intelligence for Science and Technology (QuArtist)
and Physics Department, Shanghai University, 200444 Shanghai , China\\
$^{3}$Humboldt Study Centre, ULM University, Ulm D-89069, Germany\\
$^{4}$Beijing Computational Science Research Center, Beijing 100084, China
}\date{\today}
\begin{abstract}
The polaron picture is employed to investigate and elucidate the physics of the two-qubit quantum Rabi model, which describes two identical qubits coupled to a common harmonic oscillator. This approach enables us to obtain the ground-state energy and some other simpler physical observables with high accuracy in all regimes of the coupling strengths $g$, while there is no constraint to the ration of tunneling frequency $\Omega$ and field frequency $\omega$, which is not simultaneously possible using previous methods. Besides, we also discover a new phenomenon which is not present in the one-qubit Rabi model: with the increase of coupling strength $g$, there is a transition of the ground state of the system from a multipolaron state to a bipolaron state. However, the tunneling frequency  $\Omega$ counteracts this process. Specifically, when tunneling frequency $\Omega=0$, the system always stays in the bipolaron state.
\end{abstract}
\maketitle

\section{introduction}
The quantum Rabi model (QRM)~\cite{Rabi} is one of the most simplest and basic models of non-relativistic quantum electrodynamics (QED), the basis of quantum optics~\cite{SH}. In the context of quantum optics, it describes the interaction of an atom or quantum dot with a single-mode electromagnetic field tuned to two particular levels of the atoms or quantum dot~\cite{Leibfried,Blais}. In recent years, with the continuous breakthroughs of experimental possibilities, the ultrastrong coupling regime~\cite{JBJ,TN,JDD,AFA,BPP,JQY,ARM,YTC,ABE,PFD} between the two-level atom and the  single-mode cavity field has been achieved in circuit QED, and even deep strong coupling regime~\cite{JCG,SD,PFDJ,FYT,ZCY}. Within this range, many novel physical phenomena have been discovered in these strong coupling regimes. However, the rotation wave approximation (RWA)~\cite{ETJ} for the (QRM) breaks down when these strong coupling regimes are reached. Therefore, the full QRM has to be considered and is consequently attracting renewed attention~\cite{EK,JL,TW,LAA,LAB,DB,CLZ,HWSL,CLZS,HKDL,ZHB,ZHBX}.

In this paper, we study the two-qubit Rabi model~\cite{CS,PJA,KMC,PJB,ZYY,BBM}. This model describes the coupling between two two-level atoms and a single mode light field. It has been widely used in cavity quantum eletrodynamics system~\cite{SH}, superconducting circuit quantum electrodynamic systems~\cite{LI,GT,RGL}, as well as to model systems of quantum information science~\cite{JA} and cold atom physics~\cite{EL}. For example, in a cavity QED experiment, the entangled states of atoms~\cite{FJ,DL} can be prepared by having two atoms interact with the cavity successively. In quantum information science, double qubit logic gates~\cite{SW,MBJ} and coherent storage and transmission between two qubits ~\cite{SM} are realized by optics cavities.

However, due to the counter-rotating wave terms in the full quantum Rabi model, the system is not confined to the Fock state with a fully determined boson number during it transitions from up to down state, which makes it difficult to obtain an exact solution of the model in the strong-coupling regime. Many methods have been applied to the two-qubit quantum Rabi model to obtain exact or approximate solutions, such as  Bargmann-space techniques~\cite{PJA,PJB}, and approximate methods like the adiabatic approximation~\cite{LJM}, perturbation theory~\cite{CS}, the method of extended coherent states~\cite{CQH}, the zeroth-order approximation method~\cite{LJM} and the generalized rotating-wave approximation (GRWA)~\cite{ZYY}.

Using all of the above methods, the energy spectrum, dynamical behavior and the evolution of entanglement for the two qubits could be exactly or approximately determined. However, these methods are not universally applicable to the two-qubit quantum Rabi model (or the quantum Rabi model, for that matter). By definition, RWA neglects the counter-rotating terms in the interaction, and is therefore only valid in the regime $g\ll\omega$ and $|\omega-2\Omega|\ll|\omega+2\Omega|$. The zeroth-order approximation method works only for the tunneling frequencies $\Omega$ much smaller than the field frequency, i.e. $\omega\gg\Omega$, while the coupling strength is allowed $\Omega$ are much smaller than field frequency  and the coupling strength is allowed to reach the ultrastrong regime. The GRWA works for the ultrastrong regime, $g<\omega$ , or for negative detuning, $\delta=\omega>\Omega<0$ ~\cite{ZYY}. However, it will fail in producing the correct ground state energy when the tunneling frequencies are near the field frequency or when the system is near the quantum phase transition point~\cite{MJHR,MJH,MXL}, and none of these methods give results for correlation functions and many other simpler physical observable. Therefore, a method that can be applied to the model in the whole coupling range of coupling strengths and without limiting the value of $\Omega/\omega $ becomes very urgent.

The method we apply in our study is a variational method, which allows us to obtain the ground state energy of the model and other physical observables in the whole coupling strength $g$.

The key to solving a physical model by using a variational method is the selection for an appropriate trial wave function.

In section \ref{section:Trial}, we present the basis for selecting variational wave functions for the two-qubit Rabi model. Then, in section \ref{section:GSEO}, we compare our variational results and the GRWA results with values from exact diagonalization. The results show that our method has a high precision within whole range of coupling strengths, and that there is no restriction on the value of $\Omega/\omega $. After that, \ref{section:Cross}, we discuss a new phenomenon: with the increase of the coupling strength, the ground state shows a transition from a four-polaron ground state
\begin{equation}
\label{eq:4PolaronWF}
\Psi=\frac{1}{2}(\psi_{1}|\uparrow\uparrow\rangle+\psi_{2}|
\downarrow\downarrow\rangle-\psi_{3}|\uparrow\downarrow\rangle-\psi_{4}|
\downarrow\uparrow\rangle)
\end{equation}
to a two-polaron ground state
\begin{equation}
\label{eq:TPGS}
\Psi=\frac{1}{2}(\psi_{1}|\uparrow\uparrow\rangle+\psi_{2}|
\downarrow\downarrow\rangle),
\end{equation}
when the tunneling frequency is equal to zero, the system remains in the two-polaron ground state (\ref{eq:TPGS}). In section \ref{section:QTun}, we give a diagrammatic physical explanation of this phenomenon with the help of polaron picture~\cite{YZJ,LCS,LCM}, and demonstrate the process of qubits tunneling through the light field. In order to provide some perspective, we suggest to extend this method to the multiple-qubit case, i.e. the Dicke model. Lastly, we summerize and draw the conclusions of our study in section \ref{section:Conc}.

\section{Hamiltonian and trial wave function}
\label{section:Trial}
The two-qubit quantum Rabi model consists of two artificial two-level systems coupled to a single-mode boson field. The model Hamiltonian reads (we set  $\hbar =1$)
\begin{eqnarray}
H =\omega a^{\dagger }a+\sum_{i=1,2}\left(\frac{\Omega_{i}}{2}\sigma_{x}^{i}+g_{i}\sigma_{z}^{i}\left(a^{\dagger}+a\right)\right),
\end{eqnarray}
where $\Omega_{i} \left(i =1,2\right)$ is the tunneling frequency of the $i$th qubit, $\sigma_{z}^{i}$ and $\sigma_{x}^{i}$ are Pauli matrices of the $i$th qubit, $a^{\dagger}\left(a\right)$ denotes the bosonic creation (annihilation) operator of a single boson model with frequency $\omega$, and $g_{i}$ describes the coupling strength between the cavity and the $i$th qubit.

\begin{center}
\begin{figure*}[tbp] 
 \includegraphics[width=0.67\columnwidth]{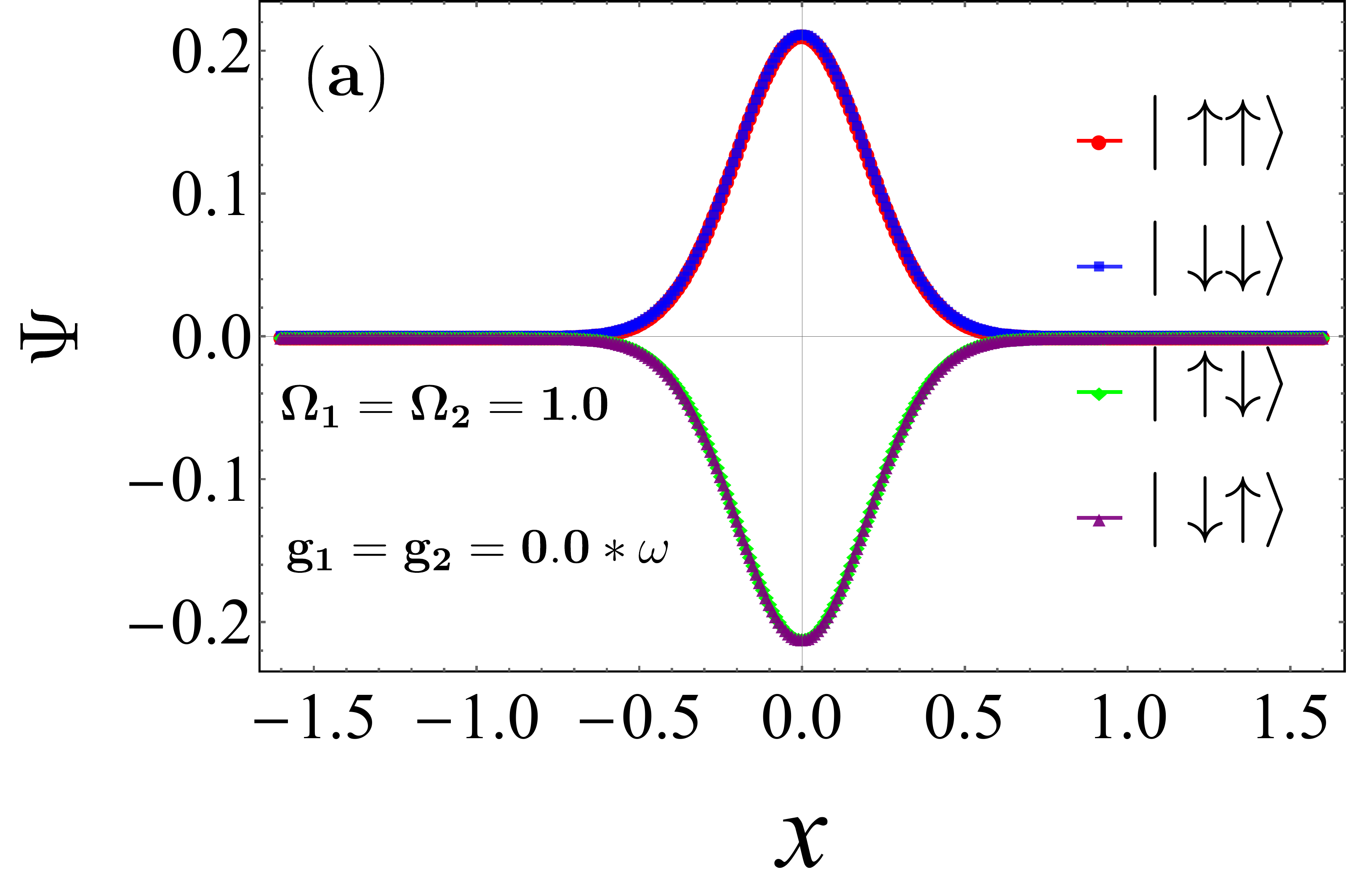}
 \includegraphics[width=0.67\columnwidth]{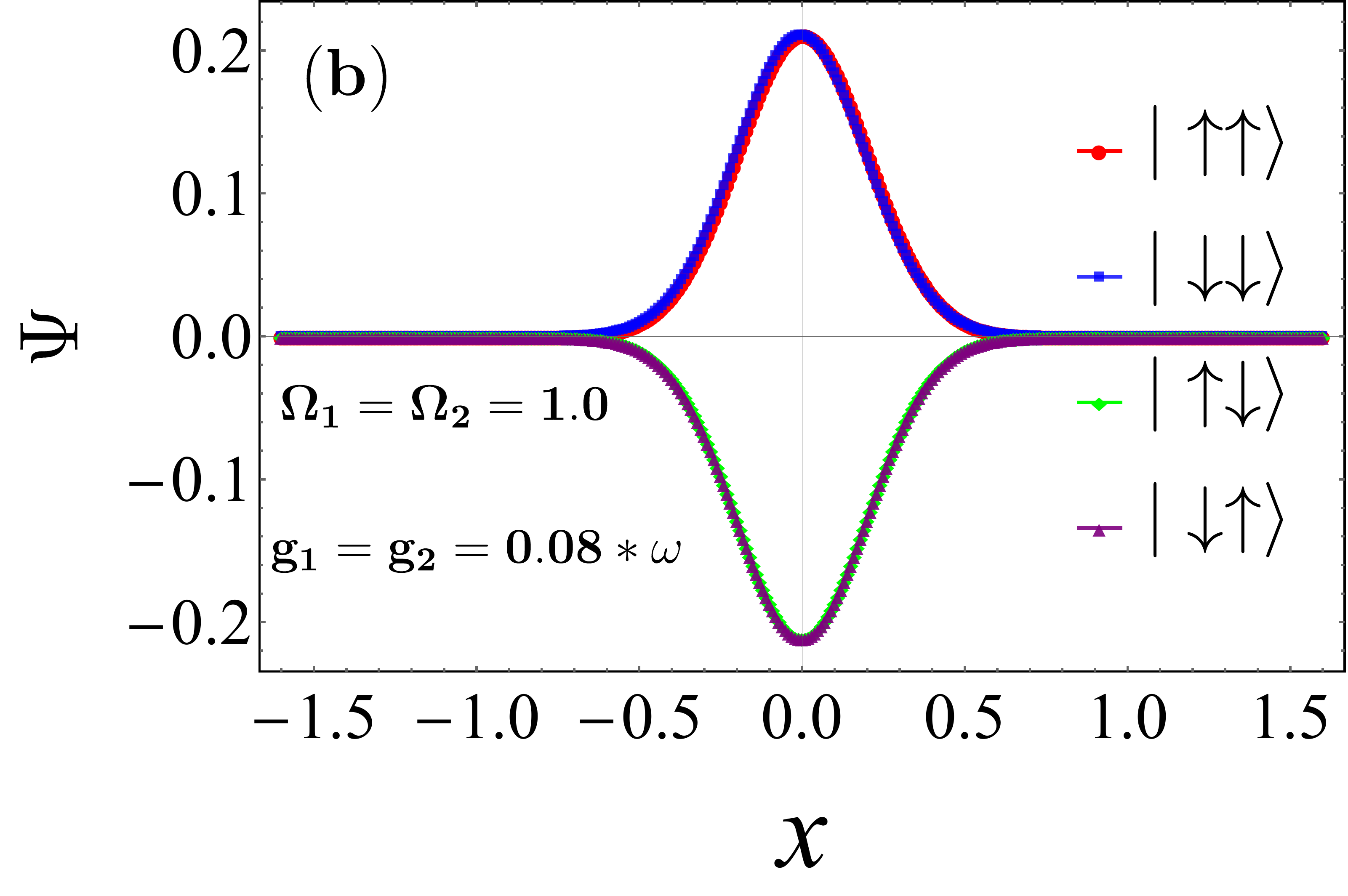}
 \includegraphics[width=0.67\columnwidth]{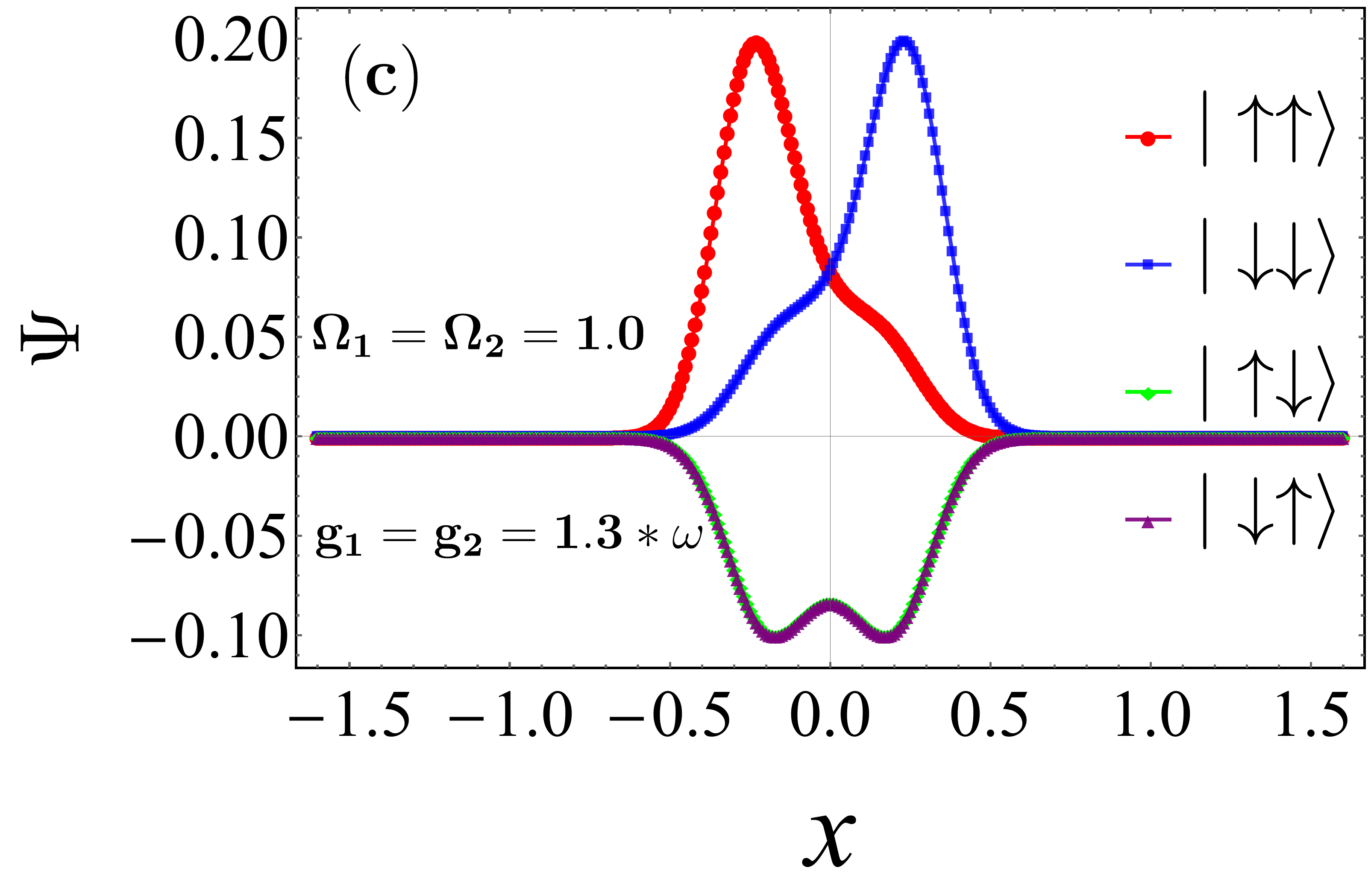}
 \includegraphics[width=0.67\columnwidth]{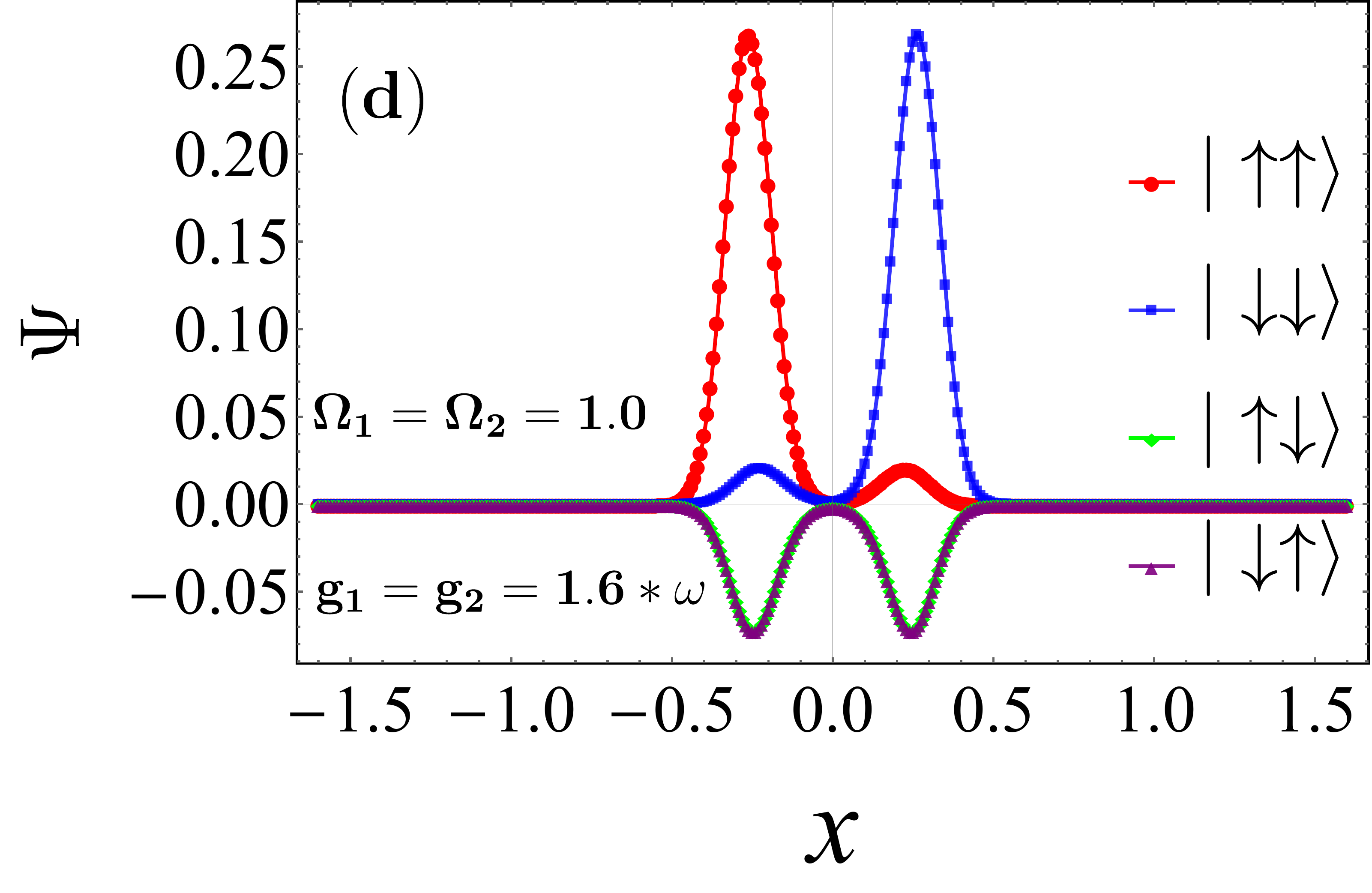}
 \includegraphics[width=0.67\columnwidth]{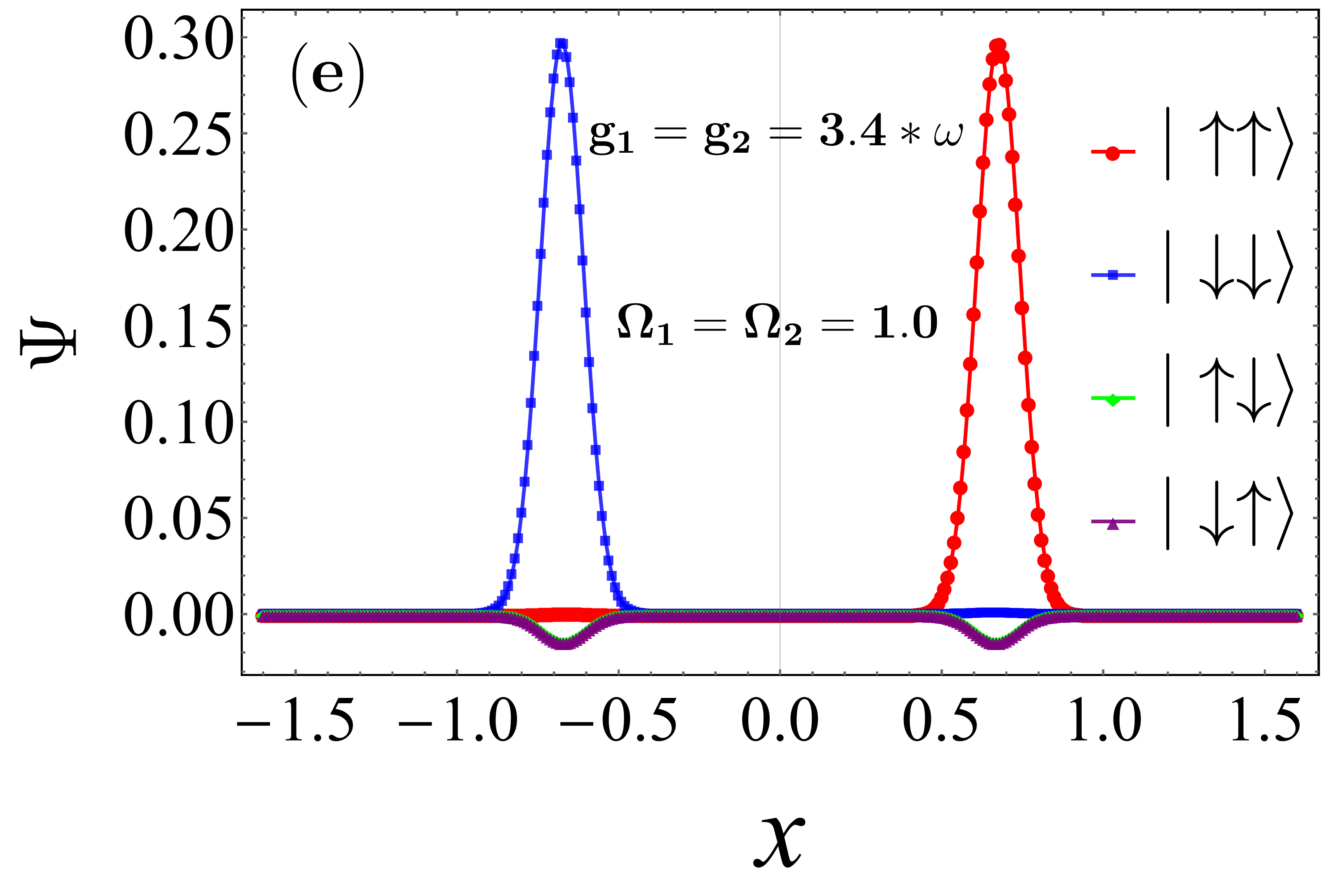}
 \includegraphics[width=0.67\columnwidth]{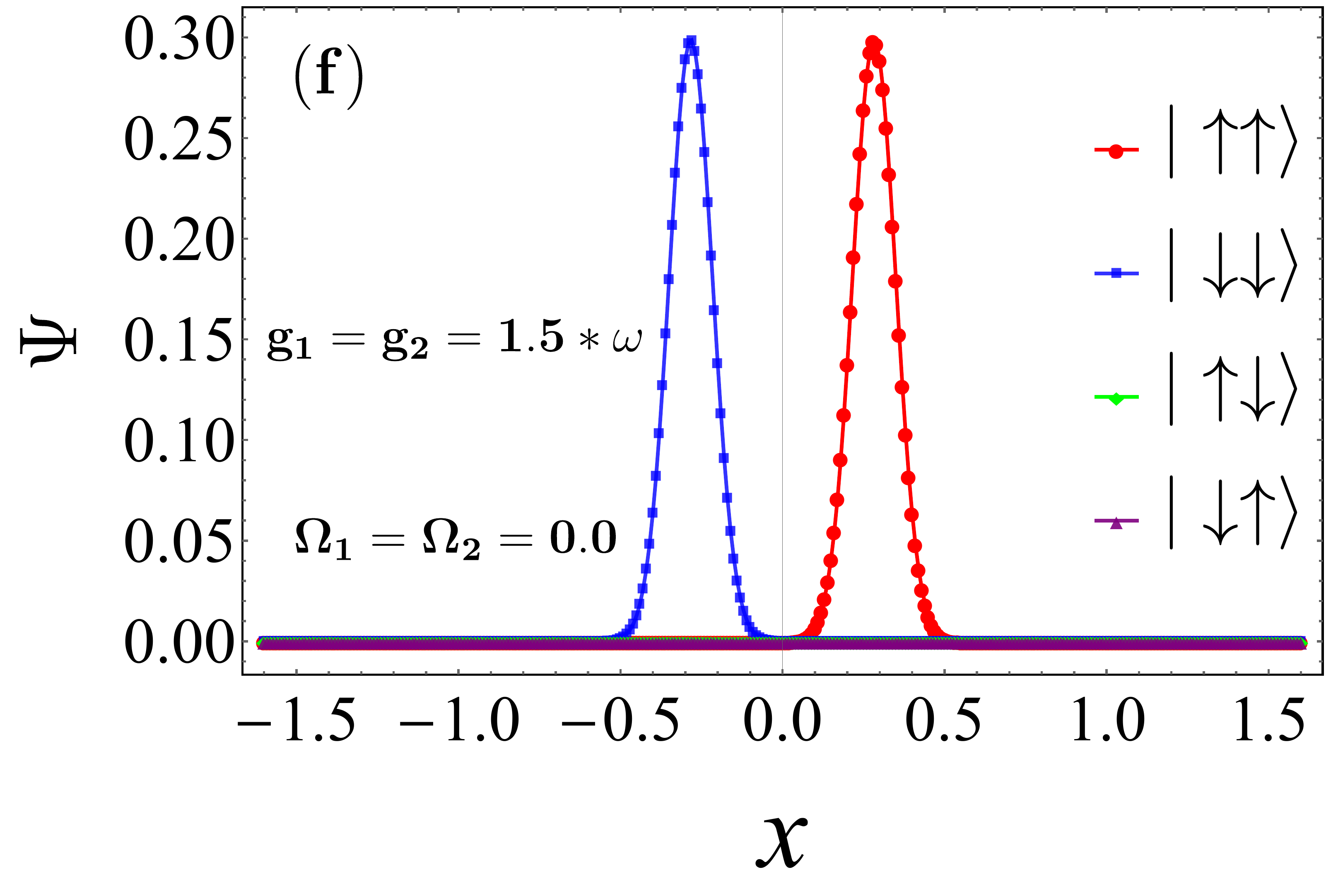}
 \caption{(Color online) The ground-state wave functions obtained by 
 numerically exact diagonalization for different $g$.
The value of the single mode frequency is $\omega=1.0$.
(a) corresponding to coupling strength $g_1=g_2=0$,
(b) corresponding to weak coupling,
(c) corresponding to strong coupling,
(e) corresponding to the ultra-strong coupling regime,
(d) corresponding to the
deep-strong coupling regime, and
(f) corresponding to tunneling frequency
$\Omega=0$, in the strong coupling regime. The red lines are the wave function component of the two qubits are in state $|\uparrow\uparrow\rangle$, the green lines represent
the wave function component the two qubits in state $|\uparrow\downarrow\rangle$, the purple lines represent the wave function component of the two qubits in state $|\downarrow\uparrow\rangle$, and finally the blue lines represent the wave function component of the two qubits in state $|\downarrow\downarrow\rangle$.}
 \label{fig:mz}
\end{figure*}
\end{center}
Denoting the position variable by $\xi$, we introduce
the dimensionless position variable $x=\xi\sqrt{m\omega}=\xi/\xi_0$.
The creation and annihilation operators of the harmonic oscillator are then given by
$a^{\dagger}=(\hat{x}-{\rm i\hat{p}})/\sqrt{2},a=(\hat{x}+{\rm i\hat{p}})/\sqrt{2}$, where $\hat{x}=x$ and $\hat{p}=-i\frac{\partial}{\partial x}$ denote the (dimesionless) position and momentum operators.
The Hamiltonian can now be rewritten as
\begin{eqnarray}  \label{crtham1}
H =\frac{\omega}{2}\left(\hat{x}^{2}+\hat{p}^{2}\right)+\sum_{i=1,2}\left(\frac{\Omega_{i}}%
{2}\sigma_{x}^{i}+\frac{g_{i}^{\prime}}{2}\omega\hat{x}\sigma_{z}^{i}\right)-\frac{\text{%
\ensuremath{\omega}}}{2},
\end{eqnarray}
where $g_{i}^{\prime}=\sqrt{2}g_{i}/\omega$.

In the following, we set $g_1=g_2$, to finally obtain the Hamiltonian
\begin{equation}  \label{eq:HH}
\begin{split}
H=&\sum_{\sigma_{z}=\pm}((h^{\sigma_{z}\sigma_{z}}-\frac{\omega}{2}g^{\prime2})|\sigma_{z}\sigma_{z}\rangle\langle\sigma_{z}\sigma_{z}|+h^{\bar{\sigma}_{z}\sigma_{z}}|\bar{\sigma}_{z}\sigma_{z}\rangle\langle\bar{\sigma}_{z}\sigma_{z}|)\\
+&\sum_{\sigma_{z}=\pm}\frac{\Omega_{1}}{2}\left(|\bar{\sigma}_{z}\sigma_{z}\rangle\langle\sigma_{z}\sigma_{z}|+|\sigma_{z}\sigma_{z}\rangle\langle\bar{\sigma}_{z}\sigma_{z}|\right)\\
+&\sum_{\sigma_{z}=\pm}\frac{\Omega_{2}}{2}\left(|\sigma_{z}\bar{\sigma}_{z}\rangle\langle\sigma_{z}\sigma_{z}|+|\sigma_{z}\sigma_{z}\rangle\langle\sigma_{z}\bar{\sigma}_{z}|\right)-\frac{\omega}{2}.
\end{split}
\end{equation}
where $\sigma_{z}=\pm$, and $\bar{\sigma}_{z}=-\sigma_{z}$, $h^{\sigma_{z}\sigma_{z}}=\frac{\omega}{2}(\hat{p}^{2}+(\hat{x}\pm g^{\prime})^{2})$, $h^{\bar{\sigma}_{z}\sigma_{z}}=\frac{\omega}{2}(\hat{p}^{2}+\hat{x}{}^{2})$, the $+\left(-\right)$ labels the up $\uparrow \left( down \downarrow\right)$ spin in the $z$ direction.

We are now going to discuss how we choose the trial wave function. Firstly, the trial wave function $\Psi$ should satisfy the Schr\"odinger equation $H\Psi=E\Psi$. Secondly, we assume that the trial wave function can be written as a direct product state such that each of the four basis state {$|\uparrow\uparrow\rangle,|\downarrow\uparrow\rangle\,|\uparrow\downarrow\rangle,|\downarrow\downarrow\rangle$} of the two-qubit system is multiplied by a wave function  $\psi_{n}$ ($n=1,2,3,4$) of the single-model optical field. In view of the fact that the two-qubit Rabi model possesses parity symmetry~\cite{LJM}, namely, $[\Pi,H]=0$, with $\Pi=\Pi_{j=1,2}\hat{\sigma_{j}^{x}exp(i\pi a^{\dagger}a)}$ the trial wave function then takes the (as anticipated in (1)) form
\begin{equation}  \label{crtham1}
\begin{split}
\Psi=&\frac{1}{2}[\psi_{1}\left(x\right)|\uparrow\uparrow\rangle+\psi_{2}\left(x\right)|\downarrow\downarrow\rangle\\
&-\psi_{3}\left(x\right)|\uparrow\downarrow\rangle-\psi_{4}\left(x\right)|\downarrow\uparrow\rangle],
\end{split}
\end{equation}
where $\psi_{1}\left(x\right)=\psi_{2}\left(-x\right),\psi_{3}\left(x\right)=\psi_{4}\left(-x\right)$.

In order to obtain expressions for the wave functions $\psi_{n}$, we analyzed the ground-state wave function for the model obtained by numerically exact diagonalization, as shown in Fig.\ 1. From the figure, we can see that, with increasing the coupling strength, the wave function changes from one Gaussian wave packet (each wave packet represents a polaron) to two Gaussian wave packets. This observation leads us to set
\begin{equation}  \label{crtham1}
\begin{split}
\psi_{n}(x)=\alpha_n\phi_{\alpha_n}(x)+\beta_n\phi_{\beta_n}(x),\quad n=1,2,3,4 \nonumber
\end{split}
\end{equation}
where one of the Gaussian wave packets is shifted to the right, the other to the left, and $\alpha_n$ and $\beta_n$  are variational parameters which we can use to adjust the trial wave function. The expressions for $\phi(x)$ take the form of Gaussian functions
\begin{equation}  \label{crtham1}
\begin{split}
\phi(x)=(\xi_{0}\sqrt{\pi})^{-\frac{1}{2}}
e^{-x^{2}/2}.
\nonumber
\end{split}
\end{equation}
When the wave function changes along with the coupling strength, our trial wave function will be able to simulate this change. The wave packet's shape and size can be varied by means of varying the frequence ($\omega\rightarrow\epsilon\omega, \xi_0\rightarrow\sqrt{\epsilon}\xi_0$) and shifting the position ($x\rightarrow\sqrt{\epsilon}(x\pm\zeta g^\prime)$).Thus, we have introduced two new parameters $\epsilon$ and $\zeta$ to adjust 
$\phi(x)$ for each of the Gaussian functions, i.e.
\begin{equation}  \label{crtham1}
\begin{split}
\phi(x)=\epsilon^{\frac{1}{4}}(\xi_{0}\sqrt{\pi})^{-\frac{1}{2}}
\exp\left(-\frac{(x\pm\zeta g^{\prime})^{2}\epsilon}{2}\right),
\nonumber
\end{split}
\end{equation}
so that we can obtain a trial wave function which is closer to the real ground state 
wave function.

We can see that our trial wave function of the two-qubit system has the same form as the trial wave function of one-qubit Rabi model based on the concept of polaron and anti-polaron which is applied to describe the phase diagram of the QRM~\cite{YZJ,LCS,LCM}. However, our trial wave function is motivated by the ground state wave function we obtained from numerically exact diagonalization. Moreover, inspired by the frequency-renormalized multipolaron expansion
method~\cite{LCS}, in order to improve the accuracy of our results, our trial wave function $\psi_{n}$ can be more generally expanded in $N$ pairs of Gaussian wave packets as
\begin{equation}  
\begin{split}
\psi_{1}\left(x\right)=\psi_{2}\left(-x\right)=&\sum_{i=1}^N\left(\alpha^{(i)}_{1}
\phi_{\alpha_{1}^{(i)}}(x)+\beta_{1}^{(i)}\phi_{\beta_{1}^{(i)}}\left(x\right)\right),\\
\psi_{3}\left(x\right)=\psi_{4}\left(-x\right)=&\sum_{i=1}^N
\left(\gamma_{3}^{(i)}\phi_{\gamma_{3}^{(i)}}(x)+
\delta^{(i)}_{3}\phi_{\delta^{(i)}_{3}}\left(x\right)\right).
\end{split}
\end{equation}
Alternatively, and I have a slight preference for this version,
we could simply write
\begin{equation}
\label{eq:Multi}
\psi_{n}\left(x\right)=\sum_{i=1}^N\left(\alpha^{(i)}_{n}
\phi_{\alpha_{n}^{(i)}}(x)+
\beta_{n}^{(i)}\phi_{\beta_{n}^{(i)}}\left(x\right)\right).
\end{equation}
In section \ref{section:GSEO} we shall choose $N$ according to the
accuracy we desire.

The average energy of the two-qubit Rabi model is
\begin{equation}  \label{VarEn}
\begin{split}
E=\langle\Psi|H|\Psi\rangle,
\end{split}
\end{equation}
subject to the normalization condition
\begin{equation}  \label{Norm}
\begin{split}
\langle\psi|\psi\rangle=\frac{1}{2}\left(\langle\psi_{1}|\psi_{1}\rangle+\langle\psi_{3}|\psi_{3}\rangle\right)=1,
\end{split}
\end{equation}
We obtain the ground state energy from the condition $\delta E=0$ for a variational extremum.

Inserting a trial wave function with an appropriately chosen
number $N$ of pairs of Gaussian wave functions and the Hamiltonian of the two-qubit Rabi model into
Eqs.\ (\ref{VarEn}) and (\ref{Norm}),
we get the expression for the energy
\begin{equation}  \label{crtham1}
\begin{split}
E=&\frac{1}{2}(\langle\psi_{1}\left(x\right)|h_{1}^{++}|\psi_{1}\left(x\right)\rangle+\langle\psi_{3}\left(x\right)|h_{2}^{+-}|\psi_{3}\left(x\right)\rangle)\\
-&\Omega\langle\psi_{1}\left(x\right)|\psi_{3}(x)\rangle\\
-&\frac{\omega}{2}{g^{\prime}}^{2}-\frac{\text{\ensuremath{\omega}}}{2}
\end{split}
\end{equation}
where we have set $\Omega_1=\Omega_2$ and
\begin{equation}  \label{crtham1}
\begin{split}
h_{1}^{++}=\frac{\omega}{2}\left(\hat{p}^{2}+\left(\hat{x}+2g^{\prime}\right)^{2}\right),\nonumber\\
h_{2}^{+-}=\frac{\omega}{2}\left(\hat{p}^{2}+\hat{x}^{2}+ 4{g^{\prime}}^{2}\right).\nonumber\\
\end{split}
\end{equation}

\section{Ground-state energy and  observables}
\label{section:GSEO}
Let us start by reminding ourselves of the problem that we described and set out to improve on at the beginning of this paper: all approximations collapse when the ratio $\Omega/\omega$ grows. To this end, we compare our results obtained by the proposed variational method with results from exact numerical diagonalization, as shown in Fig.\ 2. We consider the case of $\Omega/\omega=10$ (namely, near the phase transition
point $g_c=\sqrt{\omega\Omega/2}/2$~\cite{MXL}) and take the number of pairs of Gaussian wave packets $N=2$ in equation (\ref{eq:Multi}). We find that our variational wave function yields a high accuracy and the ground-state energy, the mean photon number $\langle a^{\dagger}a\rangle$, the tunneling strength $\langle\sigma^1_{x}+\sigma^2_{x}\rangle$, the correlation function $\ensuremath{\langle(\sigma^1_{z}+\sigma^2_{z})(a^{\dagger}+a)\rangle}$
are in good agreement with those obtained by exact numerical diagonalization.
\begin{figure}[tbp]
 \includegraphics[width=1.0\columnwidth]{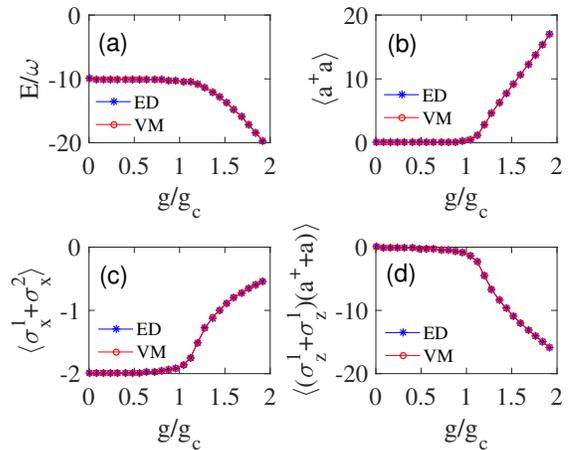}
 \caption{(Color online) Ground-state energy and some other
physical quantities as functions of the coupling strength $g/g_c$, where
$g_c=\sqrt{\omega\Omega/2}/2$. Here, $\Omega/\omega=10$.
(a) The ground-state energy.
(b) The mean photon number $\langle a^{\dagger}a\rangle$.
(c) The tunneling strength $\langle\sigma^1_{x}+\sigma^2_{x}\rangle$.
(d) The correlation function
$\ensuremath{\langle(\sigma^1_{z}+\sigma^2_{z})(a^{\dagger}+a)\rangle}$.
The blue stars denote the numerically exact results as a benchmark.
The red circles denote our results obtained by the variational method.}
 \label{fig:mz}
\end{figure}

In order to confirm that our method is better than GRWA, we made a comparison of errors of the ground state energy between the variational method and GRWA with different values of $\Omega/\omega$ in Fig.\ 3. The accuracy of GRWA, while reasonable for small values of the coupling $g$, deteriorates increasingly for $g$ approaching $g_c$ and for values of the coupling $g>g_c$. However, we find that our method works very well in all parameter regimes.

But exactly how accurate is our method? That is a question that we have been very interested in. In Fig.\ 4 we show the errors of the ground-state energy and other physical observables
obtained by the variational method as compared to results
from exact diagonalization for $\Omega/\omega=10$. We find that our variational method possesses a high degree of accuracy. For the ground state energy, the relative accuracy is as high as $10^{-4}$, and for the other physical observables, the relative accuracy is $10^{-3}$. All of these data show that our method is really very accurate, and our
variational wave function also achieves the precision we need.
\begin{figure}[tbp] 
 \includegraphics[width=1.0\columnwidth]{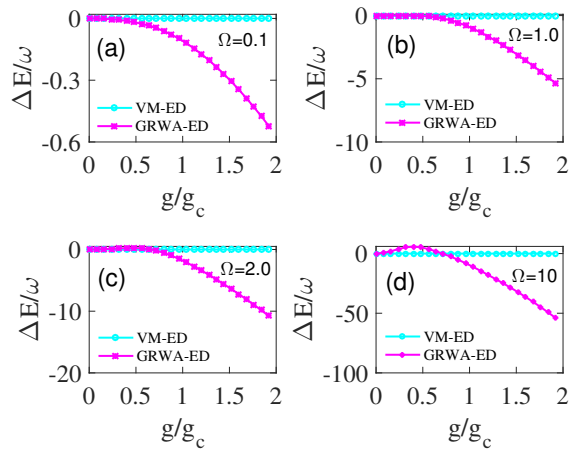}
 \caption{(Color online)  Comparison of errors of the ground state energy between our result and  the GRWA . The purple lines are the ground-state energy errors between (GRWA) and the numerically exact (ED) results. The cyan lines are the errors between our method and the ED result. Here we take  $\Omega/\omega=0.1, 1.0, 2.0, 10$ from (a) to (d)}
 \label{fig:mz}
\end{figure}
\begin{figure}[tbp] 
 \includegraphics[width=1.1\columnwidth]{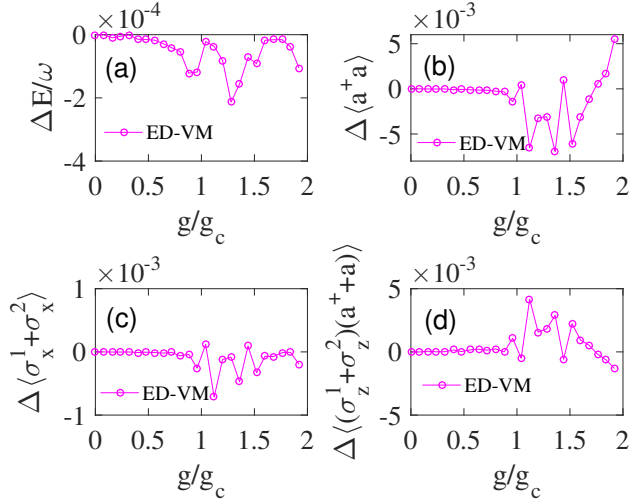}
 \caption{(Color online)  The errors of the ground state energy and other physical observables for $\Omega/\omega=10$. (a) Error in the ground-state energy with respect to the ED result
 $\Delta E=E_{\mathrm{VM}}-E_{\mathrm{ED}}$.
(b)  Error in the mean photon number $\Delta \langle a^{\dagger}a\rangle$.
(c) Error in the tunneling strength
$\Delta \langle\sigma^1_{x}+\sigma^2_{x}\rangle$.
(d) Error in the correlation function $\Delta \langle(\sigma^1_{z}+\sigma^2_{z})(a^{\dagger}+a)\rangle$. }
 \label{fig:mz}
\end{figure}
\section{Crossover from four polarons to two polarons}
\label{section:Cross}
From Fig.\ 1, we can also see there are two distinct regimes that do not appear in the one-qubit Rabi model : with increasing coupling strength $g$, the ground state wave function of the system undergoes a transitions from a multipolaron state.
\[
\Psi=\frac{1}{2}(\psi_{1}|\uparrow\uparrow\rangle+
\psi_{2}|\downarrow\downarrow\rangle-
\psi_{3}|\uparrow\downarrow\rangle-
\psi_{4}|\downarrow\uparrow\rangle)
\]
to a bipolaron state
\[\Psi=\frac{1}{2}(\psi_{1}|\uparrow\uparrow\rangle+\psi_{2}|
\downarrow\downarrow\rangle)\]

Figs.\ 1(a), 1(b), 1(c), and 1(d) clearly show the characteristics of multipolaron states while Figs.\ 1(e) and 1(f) have the characteristics of bipolaron states.
\begin{figure}[thbp]
 \includegraphics[width=1.0\columnwidth]{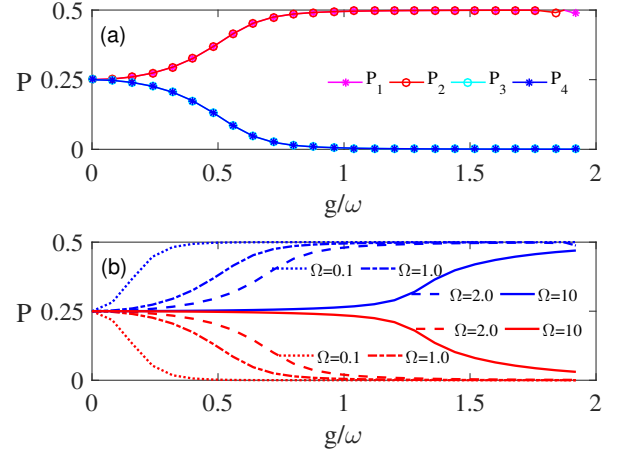}
\caption{(color online) (a) The probabilities $P_n, n=1,2,3,4$ of the components of the multipolaron state for $\Omega=1,\omega=1$. (b) The probabilities $P_1=P_2$ (blue lines) and $P_3=P_4$(red lines) for $\Omega=0.1$, $\Omega=1$, $\Omega=2$, $\Omega=10$, the single-mode frequency is $\omega=1$.}
\end{figure}
The transition from multipolaron to bipolaron state can also be seen from the probabilities $P_n=\langle\psi_n|\psi_n\rangle, (n=1,2,3,4)$ of the components of the system being in the qubit-states  $|\uparrow\uparrow\rangle$,
$|\downarrow\uparrow\rangle$, $|\uparrow\downarrow\rangle$ and $|\downarrow\downarrow\rangle$, where due to the $Z_2$ symmetry of the model $P_1=P_2$ and $P_3=P_4$.

As can be seen from Fig.\ 5(a), for $g=0$, the state of the system has equal probability $P_n=1/4\; (n=1,2,3,4)$ for all components of the four two-qubit basis states
$|\uparrow\uparrow\rangle$, $|\downarrow\downarrow\rangle$,
$|\downarrow\uparrow\rangle$, $|\uparrow\downarrow\rangle$.
With increasing coupling strength $g$, the probabilities $P_3$ and $P_4$ of the components of the multipolaron state corresponding to the two-qubit basis states
$|\uparrow\downarrow\rangle$ and $|\downarrow\uparrow\rangle$ become smaller and smaller and  finally they are reduced to zero. The probabilities $P_1$ and $P_2$ of the components corresponding to $|\uparrow\uparrow\rangle$ and $|\downarrow\downarrow\rangle$ increase gradually until they are up to $1/2$, as also shown in Fig.\ 5(b) for different values of $\Omega$. 

An increasing tunneling frequency $\Omega$ suppresses the formation of the bipolaron state. To show this effect, we calculate the probabilities $P_1$ and $P_3$ for several values of $\Omega$, displayed in Fig.\ 5(b). We can see that, when we increase the tunneling frequency $\Omega$, the transition from a multipolaron to a bipolaron state is less and less pronounced and sets in only for higher values of the coupling $g$.
\begin{figure}[thbp]
 \includegraphics[width=1.0\columnwidth]{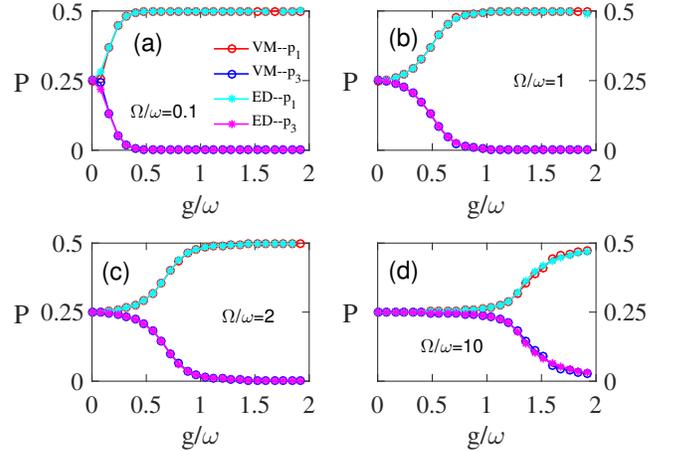}
\caption{(color online) The probabilities $P_1$ and $P_3$  for different values of $\Omega$. Here, $\omega=1$ and $\Omega=0.1, 1, 2, 10$ from (a) to (d). The purple and cyan dots denote the numerically exact results as a benchmark. The red and blue dots denote our results obtained for the variational method.}
\end{figure}

In order to further demonstrate the accuracy of our trial wave function and thus the advantages  of our variational method, we use the trial wave function obtained by the variational method to calculate the probabilities $P_1$ and $P_3$ and compare with the results obtained by exact numerical diagonalization, see Fig.\ 6. We can see that for different values of the tunneling frequency $\Omega$, the transition behavior from the multipolaron to the bipolaron state obtained  by the variational method is completely consistent with that of exact numerical diagonalization, which shows the accuracy of our trial wave function.

\section{The process of qubits tunneling through the light field}
\label{section:QTun}
\begin{center}
\begin{figure*}[tbp]
 \includegraphics[width=2.0\columnwidth]{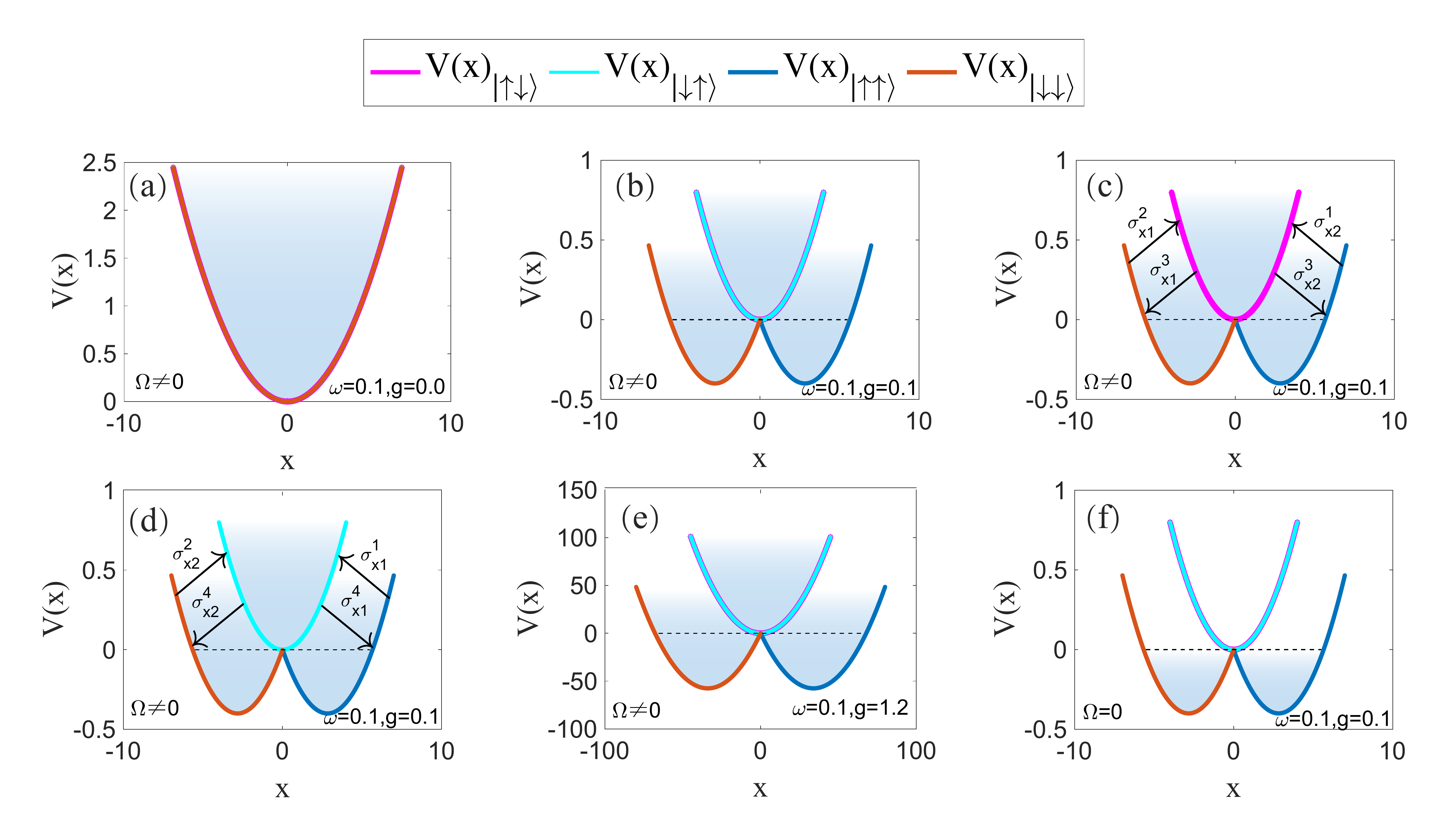}
\vspace{0.1cm}
\caption{(color online) Schematic diagram of two-qubit tunneling through the potentials created by the light field. These potentials depend on the states of the two two-lever atoms or qubits. The blue curve represents potential well $\frac{\omega}{2}(\hat{x}^2+4g'\hat{x})$ of $|\uparrow\uparrow\rangle$, the red one  represents $\frac{\omega}{2}(\hat{x}^2-4g'\hat{x})$ of $|\downarrow\downarrow\rangle$, the magenta one represents $\frac{\omega}{2}\hat{x}^{2}$ of $|\uparrow\downarrow\rangle$, and the cray one represents $\frac{\omega}{2}\hat{x}^{2}$ of $|\downarrow\uparrow\rangle$, and the blue fill indicates where the system is likely to stay in the potential wells. (a),(b),(c),(d) and (e) are for $\Omega\neq0$. (f) is in the absence of tunneling, i.e. $\Omega=0$. (a) shows the special case $g=0$, (b), (c), (d) and (e) are for $\Omega\neq0, g\neq0$. (c), (d) illustrate the process of the two-qubit atoms tunneling through the potential created by the light field.
In (e) the four potential wells are depicted for a larger coupling $g$, here for $g=1,2$. (f) When tunneling frequency $\Omega=0$, the diagram of four potential wells.}
\end{figure*}
\end{center}
For $g=0$, the Hamiltonian (\ref{eq:HH}) becomes
\begin{equation}  \label{crtham1}
\begin{split}
H=&\sum_{\sigma_{z}=\pm}\left(h^{\sigma_{z}\sigma_{z}}|\sigma_{z}\sigma_{z}\rangle\langle\sigma_{z}\sigma_{z}|+h^{\bar{\sigma}_{z}\sigma_{z}}|\bar{\sigma}_{z}\sigma_{z}\rangle\langle\bar{\sigma}_{z}\sigma_{z}|\right)\\
+&\sum_{\sigma_{z}=\pm}\frac{\Omega_{1}}{2}\left(|\bar{\sigma}_{z}\sigma_{z}\rangle\langle\sigma_{z}\sigma_{z}|+|\sigma_{z}\sigma_{z}\rangle\langle\bar{\sigma}_{z}\sigma_{z}|\right)\\
+&\sum_{\sigma_{z}=\pm}\frac{\Omega_{2}}{2}\left(|\sigma_{z}\bar{\sigma}_{z}\rangle\langle\sigma_{z}\sigma_{z}|+|\sigma_{z}\sigma_{z}\rangle\langle\sigma_{z}\bar{\sigma}_{z}|\right)-\frac{\omega}{2},
\end{split}
\end{equation}
where $h^{\sigma_{z}\sigma_{z}}=h^{\bar{\sigma}_{z}\sigma_{z}}=\frac{\omega}{2}(\hat{p}^{2}+\hat{x}{}^{2})$.

Because $h^{\sigma_{z}\sigma_{z}}=h^{\bar{\sigma}_{z}\sigma_{z}}$, the potentials~\cite{AJL,WEP} corresponding to the four two-qubit basis states
$|\uparrow\uparrow\rangle$,
$|\downarrow\uparrow\rangle$, $|\uparrow\downarrow\rangle$ and
$|\downarrow\downarrow\rangle$ are equal to $V=\frac{\omega}{2}\hat{x}^{2}$,
as shown in Fig. 7(a). Tunneling is strongest at this point, and the probabilities $P_n$ corresponding to the two-qubit basis states $|\uparrow\uparrow\rangle$, $|\downarrow\downarrow\rangle$, $|\uparrow\downarrow\rangle$ and $|\downarrow\uparrow\rangle$ are equal to $1/4$. These probabilities  of $1/4$ can be seen from Fig.\ 5 when $g=0$.

When we consider the Hamiltonian (\ref{eq:HH}) for arbitrary coupling strength $g$, we can see that the potentials depending on the four two-qubit basis states are schematically shown in the Fig. 7(b).
\begin{equation}  \label{crtham1}
\begin{split}
V_{|\downarrow\uparrow\rangle} =
V_{|\uparrow\downarrow\rangle}=&\frac{\omega}{2}\hat{x}^{2},
\nonumber\\
V_{|\uparrow\uparrow\rangle}=&\frac{\omega}{2}\left((\hat{x}-2g^{\prime})^{2}
+4g^{\prime2}\right),
\nonumber\\
V_{|\downarrow\downarrow\rangle}=&\frac{\omega}{2}\left((\hat{x}+2g^{\prime})^{2}
-4g^{\prime2}\right),
\end{split}
\end{equation}
we can see from the expression of potential $V_{|\uparrow\uparrow\rangle}$ and $V_{|\downarrow\downarrow\rangle}$ that the coupling $g$ shifts the vertices of the potential wells to the left and the right. The vertices of these two wells are also shifted downward, see the potential wells shown in blue and red in Fig.\ 7(b). The potentials $V_{|\uparrow\downarrow\rangle}$ and $V_{|\downarrow\uparrow\rangle}$ are equal and independent from $g$.

The schematic diagram shown in Figs.\  7(c) and 7(d) suggests an interpretation of the process that takes place as the coupling strength $g$ increasing: the two qubits tunnel between the potentials by spin flips brought about by a non-zero tunneling strength $\Omega\neq0$.

The first qubit tunnels with a tunneling strength of $\sigma^1_{x}$ which can be written as
\begin{equation}  \label{crtham1}
\begin{split}
\sigma^1_{x}=\sigma^1_{x1}+\sigma^1_{x2}+\sigma^1_{x3}+\sigma^1_{x4},
\nonumber
\end{split}
\end{equation}
as Fig.\ 8(a) shows, 
where $\sigma^1_{x1}=\frac{1}{4}\langle\psi_{1}(x)|\psi_{4}(x)\rangle$,
$\sigma^1_{x2}=\frac{1}{4}\langle\psi_{2}(x)|\psi_{3}(x)\rangle$,
$\sigma^1_{x3}=\frac{1}{4}\langle\psi_{3}(x)|\psi_{2}(x)\rangle$ and
$\sigma^1_{x4}=\frac{1}{4}\langle\psi_{4}(x)|\psi_{1}(x)\rangle$.
For instance, the absolute value of tunneling strength $\sigma^1_{x1}$ gives the tunneling probability through $|\uparrow\uparrow\rangle$ to $|\downarrow\uparrow\rangle$, the other probabilities have a  corresponding interpretation. Analogously the second qubit tunnels with a tunneling strength of $\sigma^2_{x}$ written in components as
\begin{equation}  \label{crtham1}
\begin{split}
\sigma^2_{x}=\sigma^2_{x1}+\sigma^2_{x2}+\sigma^2_{x3}+\sigma^2_{x4}, \nonumber
\end{split}
\end{equation}
as Fig.\ 8(b) shows, 
where
$\sigma^2_{x1}=\frac{1}{4}\langle\psi_{1}(x)|\psi_{3}(x)\rangle$,
$\sigma^2_{x2}=\frac{1}{4}\langle\psi_{2}(x)|\psi_{4}(x)\rangle$,
$\sigma^2_{x3}=\frac{1}{4}\langle\psi_{4}(x)|\psi_{2}(x)\rangle$ and
$\sigma^2_{x4}=\frac{1}{4}\langle\psi_{3}(x)|\psi_{1}(x)\rangle$.

When the coupling strength $g$ is in the weak coupling, the separation of potential wells corresponding to the four states is small, and correspondingly, the probability of tunneling for each state is maximum, as can be seen from Fig.\ 8.

\begin{figure}[thbp] 
 \includegraphics[width=1.0\columnwidth]{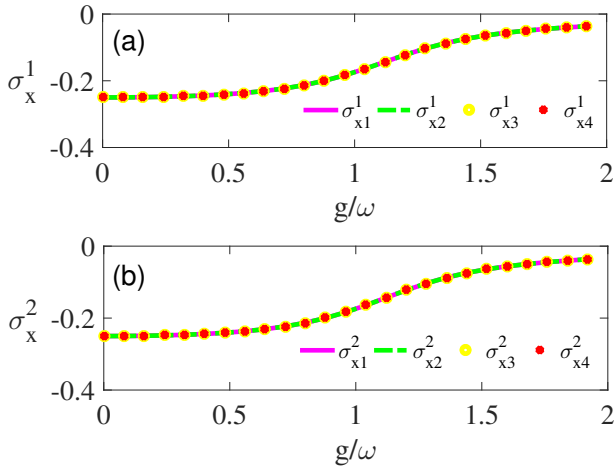}
\caption{(color online) The four components of tunneling 
strength $\sigma^1_{x}$ (a) and $\sigma^2_{x}$ (b) varying with the coupling strength $g$. Here, $\omega=1.0$, $\Omega=1.0$. \\}
 \label{fig:mz}
\end{figure}

As the coupling strength increasing, the shift of the blue and red wells to the right and left respectively increasing, and the wells become deeper. Thus, the trapping capacity of these two wells is also strengthened, as shown in Fig.\ 7(e). As a result, the probability of the two atoms staying in these two potential wells increases gradually, and the probability of the two atoms staying in the middle potential well decreases gradually. This phenomenon, reflected in the wave function, is shown in figure Fig.\ 1(b) to Fig.\ 1(d), correspondingly, the probability of tunneling for each state decrease,  as can be seen from Fig.\ 8.

When the deep strong coupling region is reached, the blue and red  potential wells will be far apart from each other, and the wells become deeper and deeper. The two atoms then stay in the red and blue wells, respectively. This is exhibited in the wave function as shown in Fig.\ 1(e).

When the tunneling frequency $\Omega=0$, the Hamiltonian of this model will be
\begin{equation}  \label{crtham1}
\begin{split}
H=&\sum_{\sigma_{z}=\pm}(h^{\sigma_{z}\sigma_{z}}-
2\omega g^{\prime2})|\sigma_{z}\sigma_{z}\rangle\langle\sigma_{z}\sigma_{z}|\\
+&\sum_{\sigma_{z}=\pm}h^{\bar{\sigma}_{z}\sigma_{z}}|\bar{\sigma}_{z}\sigma_{z}\rangle\langle\bar{\sigma}_{z}\sigma_{z}|-\frac{\omega}{2}.
\end{split}
\end{equation}
As already mentioned above, the schematic diagram of this case is shown in Fig. 7(f). Since the tunneling frequency is zero, the two atoms stay in the blue and red potential wells, respectively, which also satisfy Hund's Rule. The probabilities that the two atoms stay in $|\uparrow\downarrow\rangle$ and $|\downarrow\uparrow\rangle$, respectively, are equal to zero, and the corresponding wave function is shown in Fig.\ 1(f).

\section{CONCLUSION}
\label{section:Conc}
In summary, we demonstrated that the variational method that we apply to solve the two-qubit Rabi model has the following advantages. Firstly, both for the coupling strength $g$ and the ratio of
$\Omega/\omega$ varying over all regimes, we always achieve high accuracy, especially near the phase transition point $g_c$, under the condition of $\Omega/\omega\gg1$ the variational method can still be successfully applied to the model.

Secondly, the highly accurate values of physical observables we achieve with our variational method indicate that our trial wave function is very close to the real ground-state wave function.

Last but not least, the polaron picture can provide a helpful understanding of the transition of the ground state from the multipolaron state to the bipolaron state: as the coupling strength increasing, the potential wells of state $|\uparrow\uparrow\rangle$ and $|\downarrow\downarrow\rangle$ will become deeper, therefore the trapping capacity of the two potential wells will be strengthened, and the probability of the system to stay in $|\uparrow\downarrow\rangle$ and $|\downarrow\uparrow\rangle$ will decrease until it reaches zero, namely, the system  transition from a multipolaron state to a bipolaron state.  Meanwhile, the tunneling with frequency $\Omega$ will counteract this process, because when $\Omega$ grows, the attenuation of the tunneling strength due to the increase of the coupling strength $g$ will be reduced, that's because the probability  of staying in $|\uparrow\downarrow\rangle$ and $|\downarrow\uparrow\rangle$ is induced by the tunneling.

Moreover, we expect that our approach can be extended to the multiple-qubit case, such as the Dicke model.

\acknowledgments

We acknowledge useful discussion with Yu-Yu Zhang, Li-Wei Duan, Bin-Bin Mao, Si-Yuan Bai. This work was supported by NSFC (Grants No. 11674139 and  No. 11834005). L.C. acknowledges support by the China Postdoctoral Science Foundation (No. 2019M651463).

\end{document}